\documentclass[amsmath,amssymb,lengthcheck,aps,prl] {revtex4}
\usepackage[T1]{fontenc}
\usepackage[utf8]{inputenc}
\usepackage{graphics}
\usepackage{graphicx}

\usepackage{xcolor}
\begin{document}
 %\begin{frontmatter}
  \title{Two dimensional interacting Bose-Bose droplet in random repulsive potential}

\author{Saswata Sahu}
 \author{ Dwipesh Majumder }
\affiliation{Department of Physics, Indian Institute of Engineering Science and Technology, Shibpur, W B, India}
\date{\today}

\begin{abstract}
 We have studied the effect of time-independent repulsive random impurity potential on the quantum droplets of Bose-Einstein condensation of two different species of Bose atoms in two dimensions. We have solved the Gross-Pitaevskii equation to get the density profile and the energy of the condensate. In our study, we found that quantum droplets are incredibly robust against the non-chemical reactive impurity.
\end{abstract}
\maketitle
\section{Introduction}

%droplets
Liquid formation of the dilute ultracold atomic system \cite{Petrov2015,Trarruell2018}  is one of the most exciting topics in Bose-Einstein condensation (BEC).
The droplets have been observed in the isotropic short-range interacting system of two species of cold atoms \cite{Trarruell2018,drop_exp2, Trarruell2018PRL} as well as in the anisotropic long-range dipolar interacting system of $^{164}$Dy or $^{166}$Er atoms \cite{dipolar_droplets}. 
In the mixture of two component Bose atoms, the spherical droplet has been observed under the competition between the effective short range attractive interaction and the repulsive interaction due to the quantum fluctuation \cite{LHY}.
The two component BEC may be the mixture of atoms of two different elements (different atomic mass) \cite{PRL89, PRL100,Itali2020} or maybe the mixture of atoms with two different internal degrees of freedom of a given element \cite{spin_BEC, PRL'101, 2Comp_BEC}. 
 In the dipolar system, the cigar-shaped droplet has been observed in the balance of attractive interaction due to the asymmetric dipolar interaction and repulsive interaction due to the quantum fluctuations.  Three-body collisions limit the lifetime of the droplets. In the lower dimensions, it is expected that this lifetime can be extended because of reduced phase-space available to colliding atoms. That is why people have an interest in droplets in the lower dimension.
There are already theoretical proposals of liquid states of BEC in the lower dimensions \cite{2D_liquid,2D_liquid1, 2D_liquid2,2D_liquid3, Gajda19, SS2020}.
%%%%%%%%%%%%%%%%%%%%%%%%%%%%%%%%%%%%%%%%%%%%%%%%%%%%%%%%%%%%%%%%%%%%%%%%%%%
%impurity

	 The study of systems in the presence of impurity potential is very important and common in condensed matter physics. In most of the cases, the change of the properties of the system become very dramatic, even includes the phase transition. Usually,  in condensed matter physics, we deal with the electron, which is a fermion, but some systems such as superconductivity in which the quasiparticles are bosons, where the random potential has been included in the study \cite{Trivedi1991}. There are some studies in disordered Bose system of liquid $^4$He absorbed in various types of porous media \cite{He4porous}. 
%In the ultracold Bose atomic system, we have BEC  

	 Theoretically, the BEC in the presence of random impurity potential has been studied by Huang \cite{Huang1992} and Giorgini et al \cite{Stringari1994} before the experimental observation of BEC. 
	 The study of BEC in the presence of speckle potential \cite{speckle1, speckle2, speckle3, speckle4,speckle5, machin_learning} is very popular as the potential is controllable and random in nature. Besides the speckle potential different kind of disordered BEC has been studied, such as Gaussian random potential \cite{Gaussian1, Gaussian_Adhikari, GP_random2, Gaussian2014,Gaussian2018}, Lorentzian potential\cite{Lorentzian}.

%The experimental control over the cold atomic system is extremely precise, however, we may introduce some external noise using a signal.
Random disorder in the Bose system has many aspects, such as Anderson localization \cite{Anderson2007, Anderson2010, Anderson2008, Anderson2021}, superfluid-insulator transition  \cite{SIT3_2004, SIT1_2014, SIT2_2013}, breakdown of BEC, dynamics of impurities\cite{dynamics_imp1}, application in machine learning\cite{machin_learning}, fluctuations and superfluidity\cite{BBM-2020}. 
	 Sometimes the impurity potential enhances the confinement; even in the absence of external confinement, the impurity can bound them\cite{boundImp1}. 
	 Based on the above reference, it is a question whether the impurity will enhance the stability of the quantum droplet.

	 Disordered Bose system has been theoretically studied using different types of Monte-Carlo (MC) method \cite{PIM_2010_Pilati,diff_mc,quantum_mc}, by solving Gross-Pitaevskii (GP) equation \cite{GP_random1, GP_random2, GP_random3}, perturbative method \cite{analytical_ran1, MFT1}. 

The formation of droplets is a relatively new field, it is an important topic to see the effect of disorder potential on BEC-droplets. In this article, we have studied the effect of random external potential on the droplets of two-different species of atoms in two dimensional system by solving the GP-equation. In our study, we have chosen a suitable mathematical form of random potential, which is experimentally realizable using laser fields, such as speckle potential.

	 %Experimentally BEC with some random potential has been studied by et al. where the potential is generated by Laser. 

%The dipolar droplet in the presence of fermionic impurity has been studied by Wenzel et al. \cite{Pfau_disorder1}, 

%%%%%%%%%%%%%%%%%%%%%%%%%%%%%%%%%%%%%%%%%%%%%%%%%%%%%%%%%%%%%%%%%%%%%%%%%%%%%%%%%%
\section{System \& Method of Calculations}
\begin{figure}
  \begin{center}
    \includegraphics[width= 3.8cm]{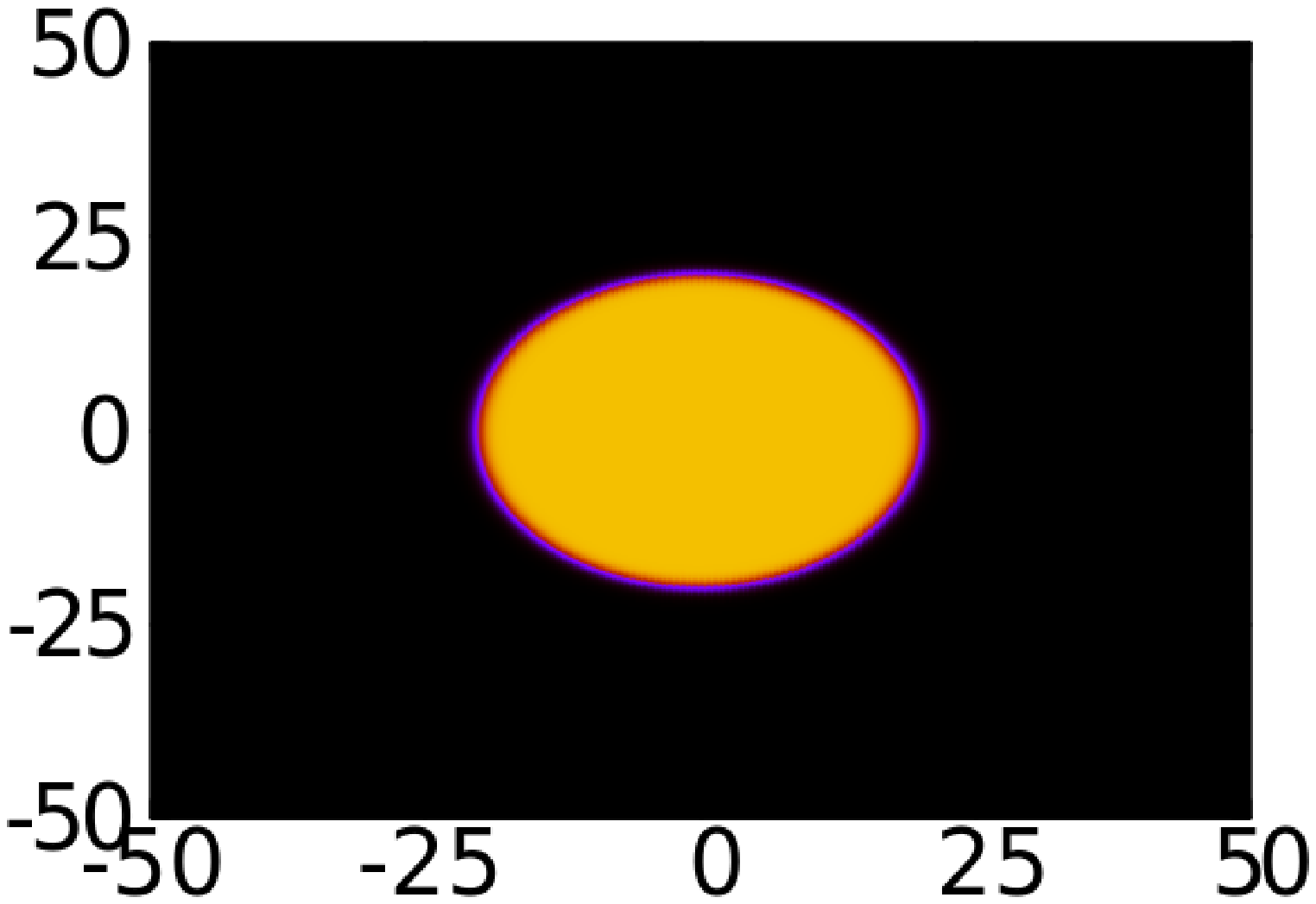}  
    \includegraphics[height=2.7cm]{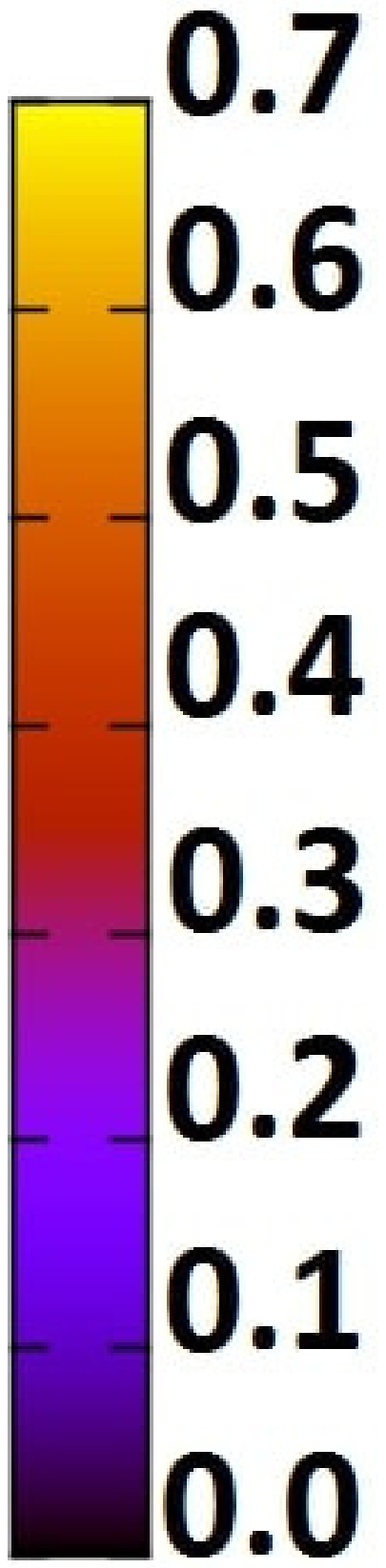} \\
    \includegraphics[width=3.6cm]{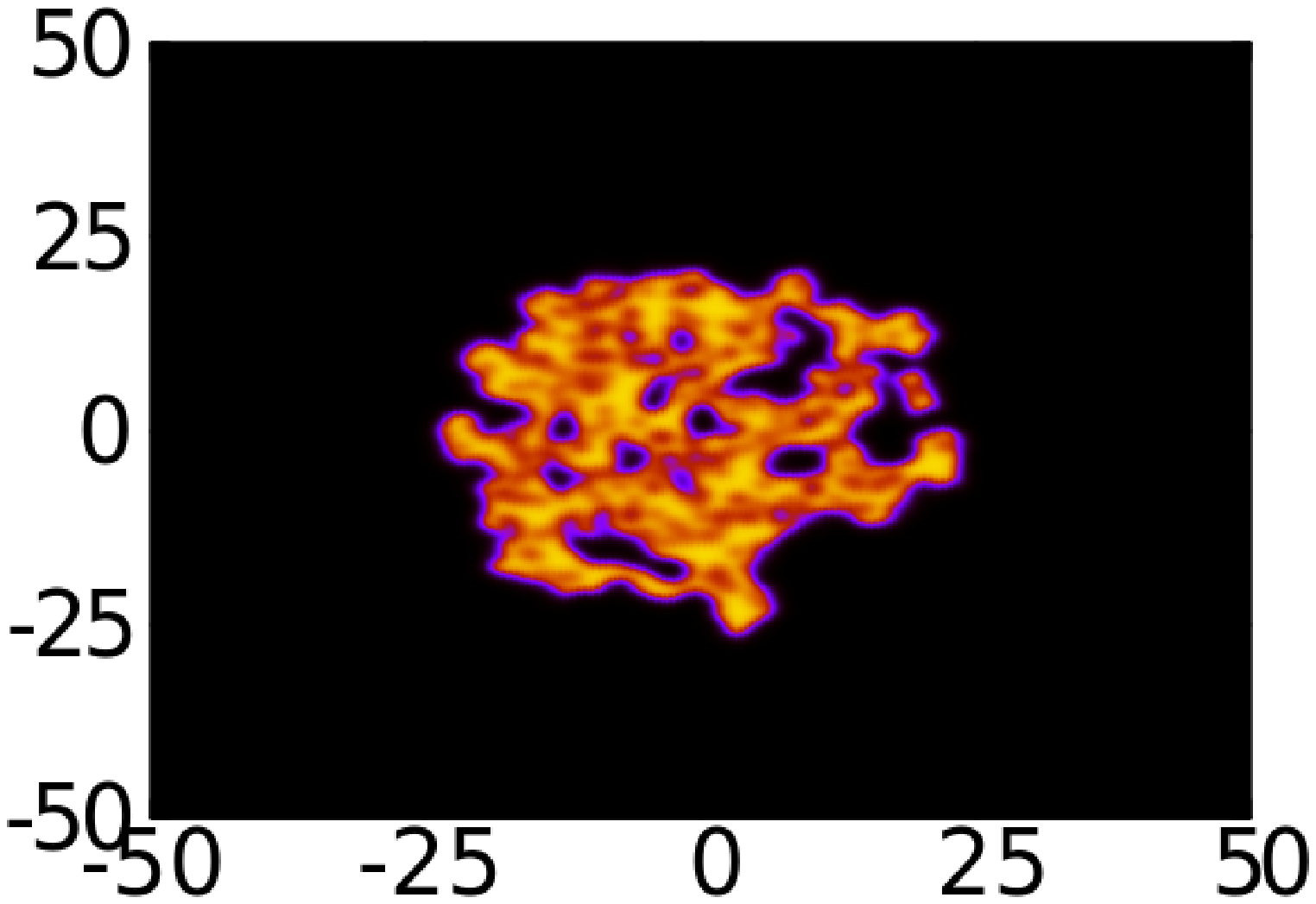} 
    \includegraphics[width=3.6cm]{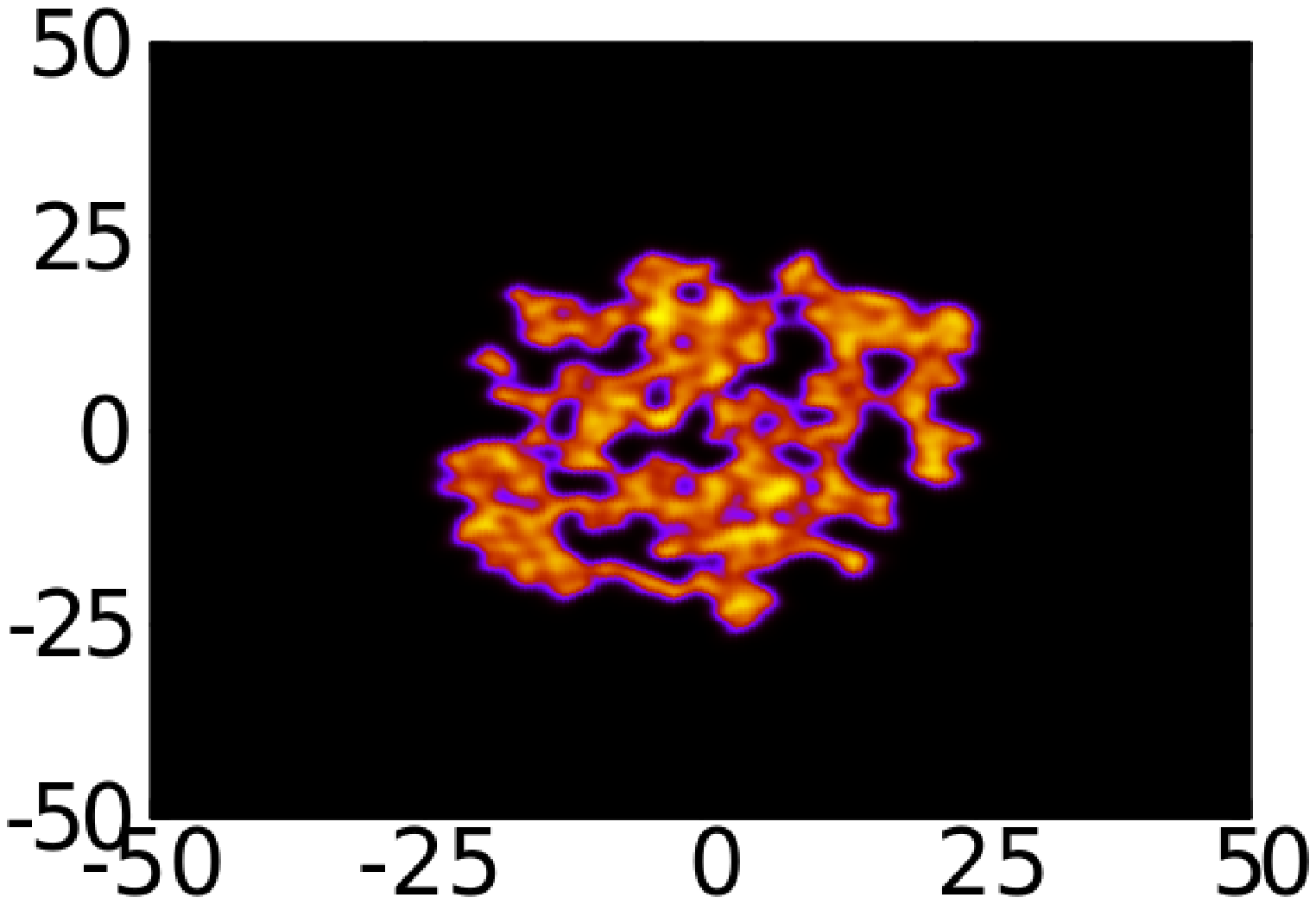} 
        \includegraphics[width=3.6cm]{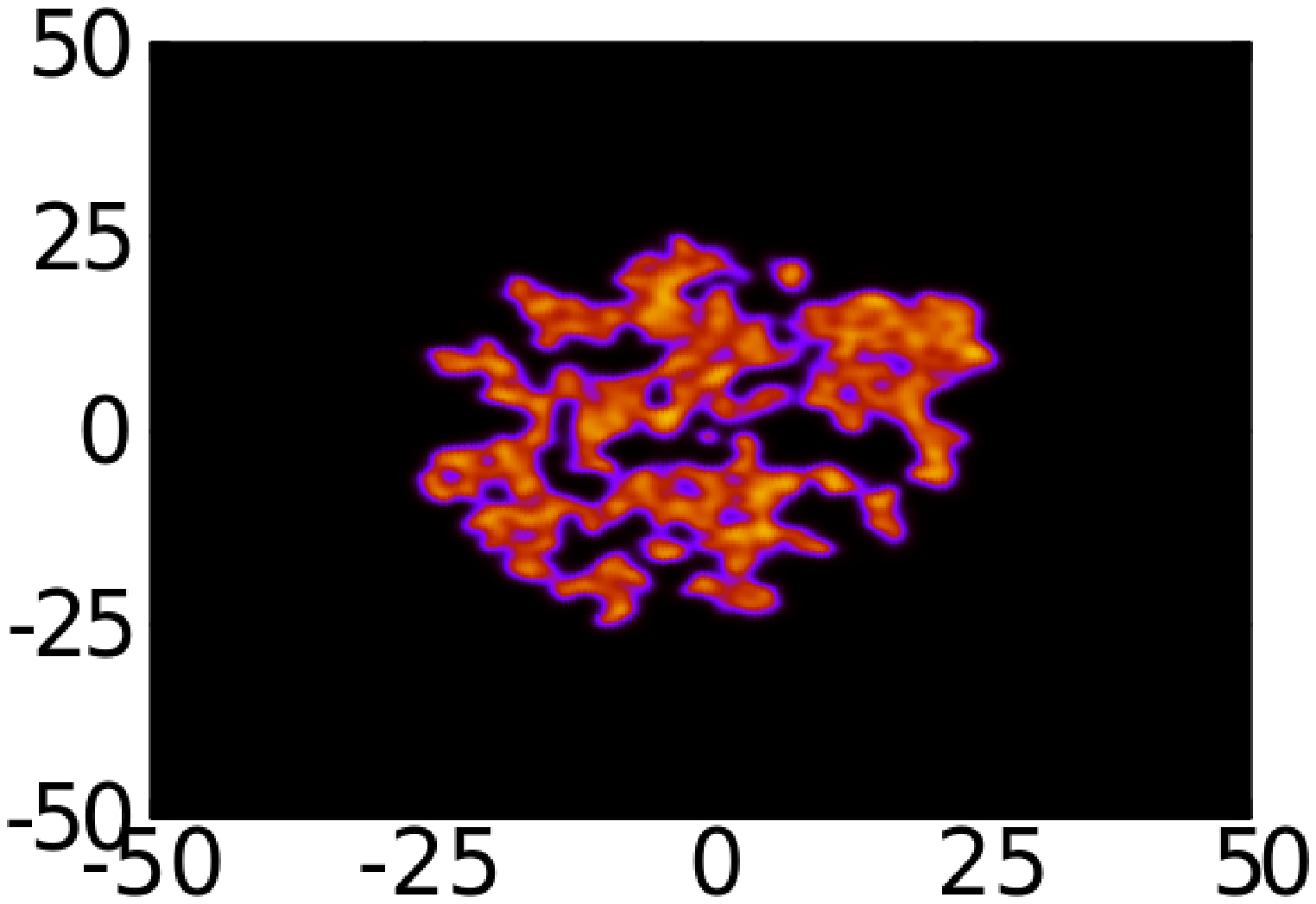} 
            \includegraphics[width=3.6cm]{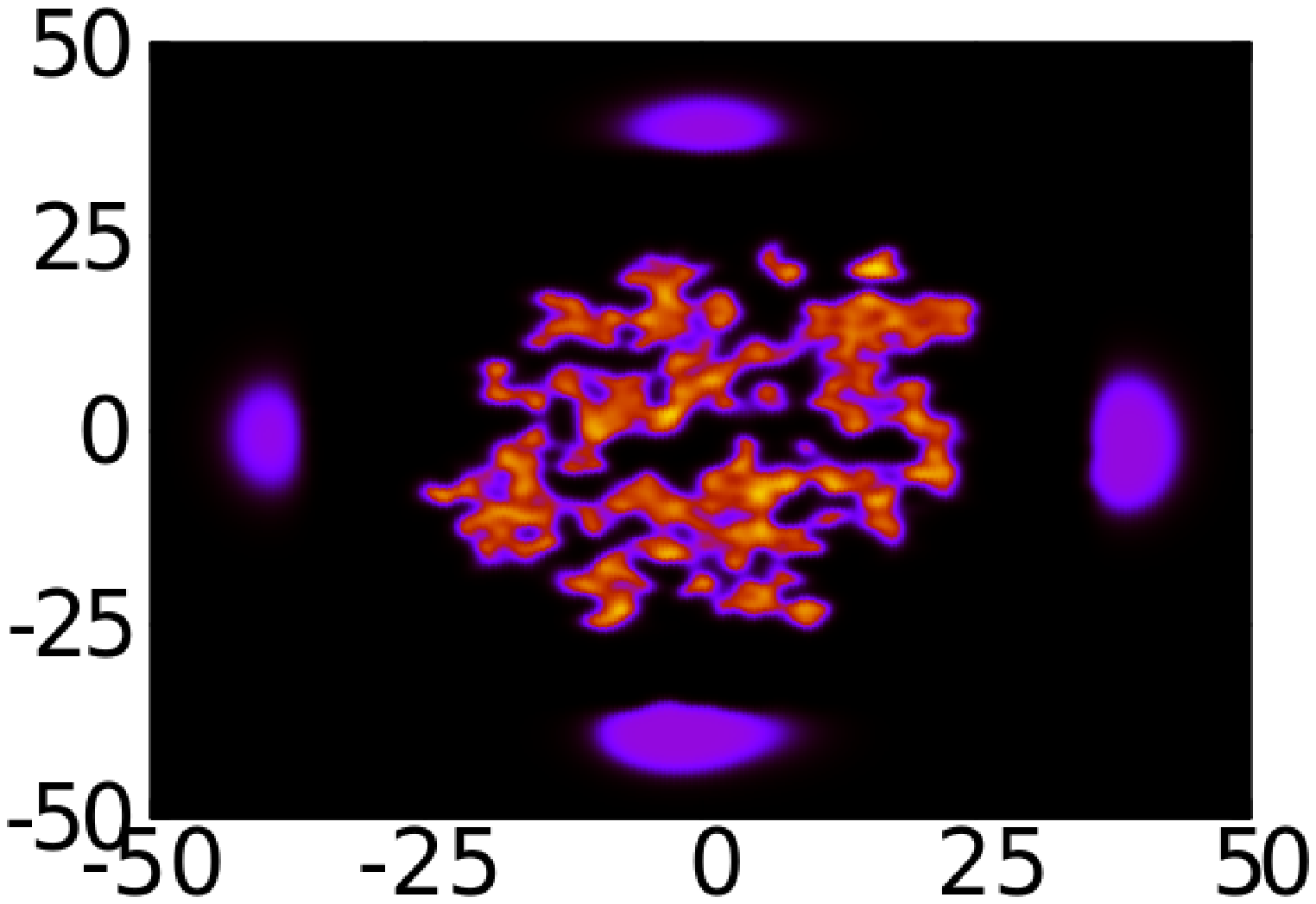} 
            \includegraphics[width=3.6cm]{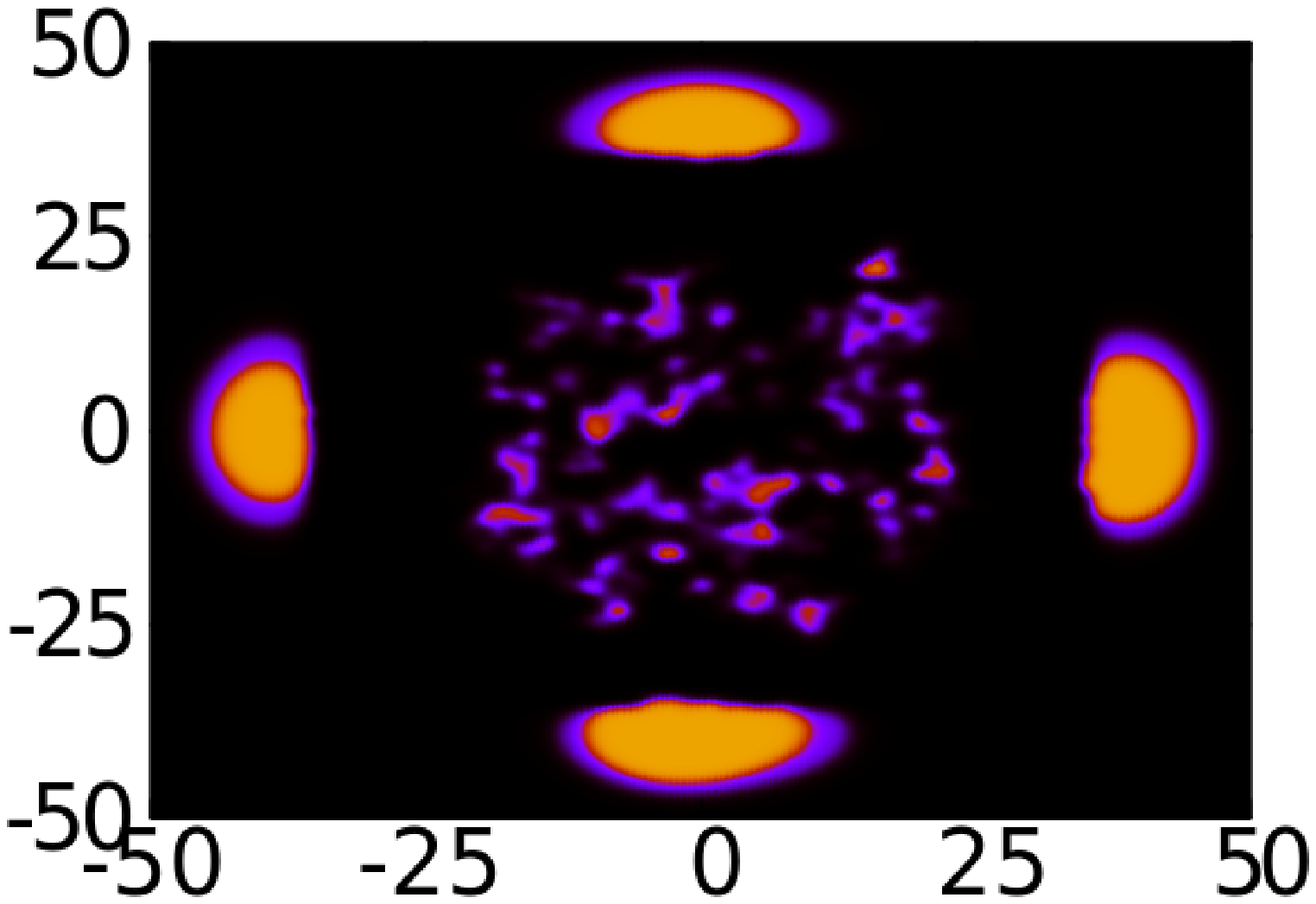} 
                \includegraphics[width=3.6cm]{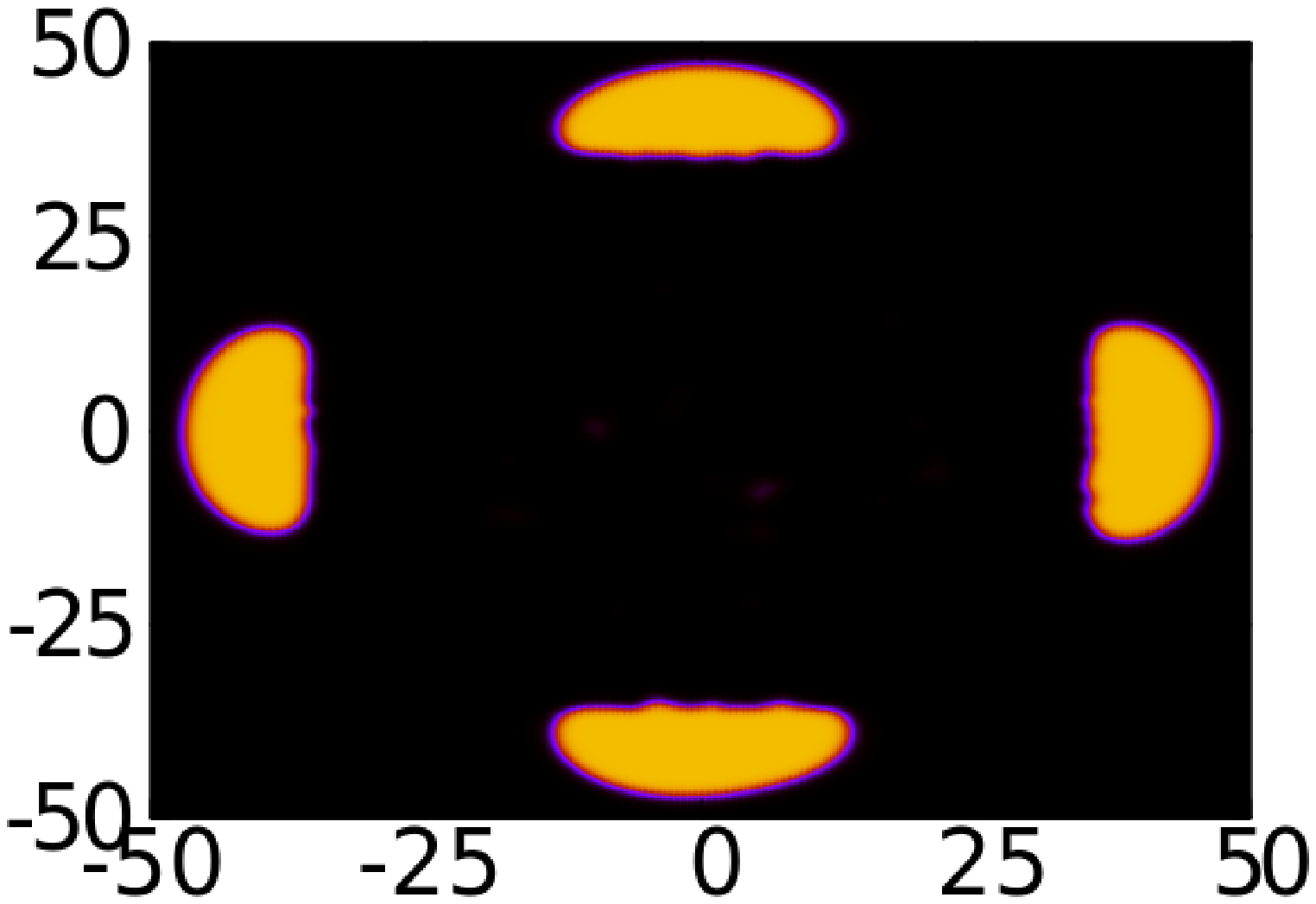} 
\caption{ \textbf{Density of the condensate}: The topmost figure is the plot of the density of the condensate in the absence of impurity potential. The next three lines are the variation of the density of the condensate for different numbers of impurity concentrations 0.408, 0.816, 1.224, 1.632, 1.836 and  2.040  in the text sequence of a system with $N=400$ number of Bose particles and the interaction between particles $g=10$.}
    \label{fig:2Dg10den}
  \end{center}
\end{figure}

\begin{figure}
  \begin{center}
 %\includegraphics[width=0.25\textwidth]{3d_rand_2000.eps}
 %\hskip -.5cm \includegraphics[width=0.25\textwidth]{3d_plot_2000.eps}   
%	  \vskip .25cm
 %\rule{0.3\textwidth}{0.4pt}
%	  \vskip .25cm
 %\includegraphics[width=0.25\textwidth]{3d_rand_6000.eps} 
 %\hskip -0.6cm \includegraphics[width=0.25\textwidth]{3d_plot_6000.eps}   
\includegraphics[width=0.55\textwidth]{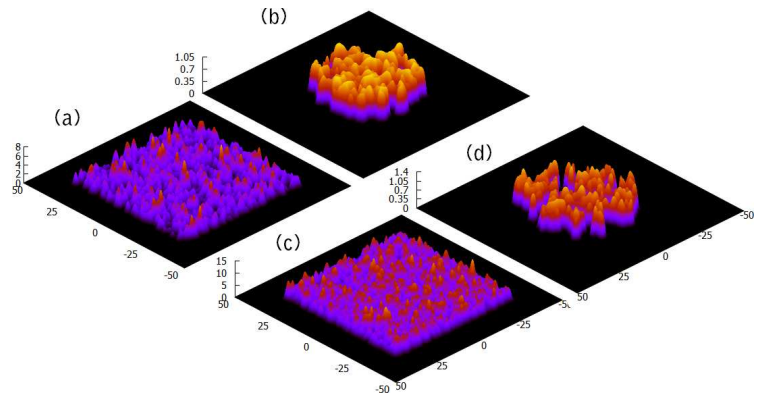}
	  \caption{ (a) Impurity potential $V(\vec{r})$ for $\rho_i = 0.408$ impurity concentration (see equation (2)). (b) The density of the condensate with the same impurity potential as that of (a).  (c) Impurity potential $V(\vec{r})$ for $\rho_i= 1.244$. (d) Density of the condensate with the same impurity potential as that of (c). Here we have considered the system size $N=400$ and interaction parameter between particles $g=10$.}
    \label{fig:2Dpotl}
  \end{center}
\end{figure}

%%%%%%%%%%%%%%%%%%%%%%%%%%%%%%%%%%%%%%%%%%%%%%%%%%%%%%%%%%%%%%%%%%%%%%%%%%%%%%%%%%
	 The mean filed Gross-Pitaevskii (GP) equation is not sufficient for the droplet; we need to consider the higher-order correction, popularly known as LHY \cite{LHY} correction term.
The well established coupled GP equations for the 2D droplets is given by \cite{2D_liquid}
% GP equations
\small{
\begin{eqnarray}
   i\frac{\partial\psi_1 }{\partial t}=\left[-\frac{\nabla ^2}{2} + g(|\psi_1|^2 - |\psi_2|^2) + \frac{g^2}{4\pi} \; \rho ln(\rho) + V(\vec{r}) \right ]\psi_1 \nonumber \\
      i\frac{\partial\psi_2 }{\partial t}=\left[-\frac{\nabla ^2}{2} + g(|\psi_2|^2 - |\psi_1|^2) + \frac{g^2}{4\pi} \; \rho ln(\rho) + V(\vec{r}) \right ]\psi_2.
\end{eqnarray}
}
\normalsize
here $V(\vec{r})$ is the external random repulsive potential.
The first term of the right-hand side is the kinetic energy term; the second is due to the mean-field part of the contact interaction; the third is the LHY correction term. Here we have considered the equal strength of repulsion between the atoms of the same species and attraction between the atoms of different species. All the quantities are expressed in suitable natural units of
the system \cite{dimension}, in which we have the unit of length $l_0$ (we have chosen $l_0=1.0\mu $m, the order of coherence length of typical BEC), the unit of energy $\hbar^2 /(m l_0^2$) ($m$ is the mass of a particle) and unit of time $ml_0^2/\hbar$.
Here we have not included any confinement potential; the attraction between different types of atoms and repulsion between the same type of atoms confined the system. The system is in the liquid phase. In our study, we have not included the three body interaction among the atoms\cite{dimension, 3body2} to avoid numerical complicacy, as two body interaction is sufficient to describe the droplet formation, and moreover, our system is sufficiently low density to avoid the three body interaction.
 We have considered Gaussian potential with random impurities points       \cite{Gaussian_Adhikari, Gaussian2018, Gaussian2014}
 \begin{equation}
   V (\vec{r}) = V_0 \sum_{i = 1}^M  e^{(-(x-x_i)^2-(y-y_i)^2)/\xi^2} 
 \end{equation}
 $V_0$ is the strength of the random potential (we have taken $V_0=1.0$ in our calculation), $\xi$ is the characterstics length of impurity potential (we have chosen $\xi=1.0$ in our calculation). We have chosen $M$ number of impurity points within the given rectangular area ($L_x \times L_y$) to get the potential, the density of the impurity points is $\rho_i = M/(L_x \times L_y)$. We allow the liquid to move out of the impurity region by setting a wall away from the impurity region. Here we have presented the results for the impurity region with area $70\times 70$ and the allowed region of the liquid with area $100\times 100$ (FIG. 2). The energy per particle does not depend on the size of the free space (without impurity region). We put the condensate at the center of the impurity potential region. The impurity points \{($x_i,y_i$)\} have been taken randomly. This type of potential can be realized using laser field to perform experiments\cite{speckle1}. The wave functions follow the normalization condition
% Normalization
 \begin{eqnarray}
   \int \left ( |\psi_1|^2 + |\psi_2|^2 \right ) d^2\vec{r} = \int \rho(\vec{r}) d^2\vec{r} = N,
 \end{eqnarray}
 %method of solution
 where $N$ is the total number of Bose particles in the condensate.
 We have used semi-implicit Crank-Nicolson (CN) method to solve the coupled GP equations. Alternative direction implicit method \cite{ADC_method} has been used to separate the x-axis and y-axis  derivative in the 2D coupled GP equation. 
 After sufficient times of iteration ($\sim 10^7 $), we have the ground state wave function, so basically we have studied stationary problem by solving the time dependent GP equation.
The calculated density of the condensate has been shown in FIG \ref{fig:2Dg10den} for the system with $g=10$.
In FIG \ref{fig:2Dpotl}, we have plotted the potential for the number density of impurity points $\rho_i = 0.408$ and $\rho_i=1.224$ the corresponding density of the condensate for the system with $g=10$ and $N = 400$.

% figure 1. Density of the condensate g=10, N=400
% figure 2, potential and corresponding density

% energy per particle and chemical potential for g=10 and g=20
\begin{figure}
  \begin{center}
	   \includegraphics[width=9.5cm]{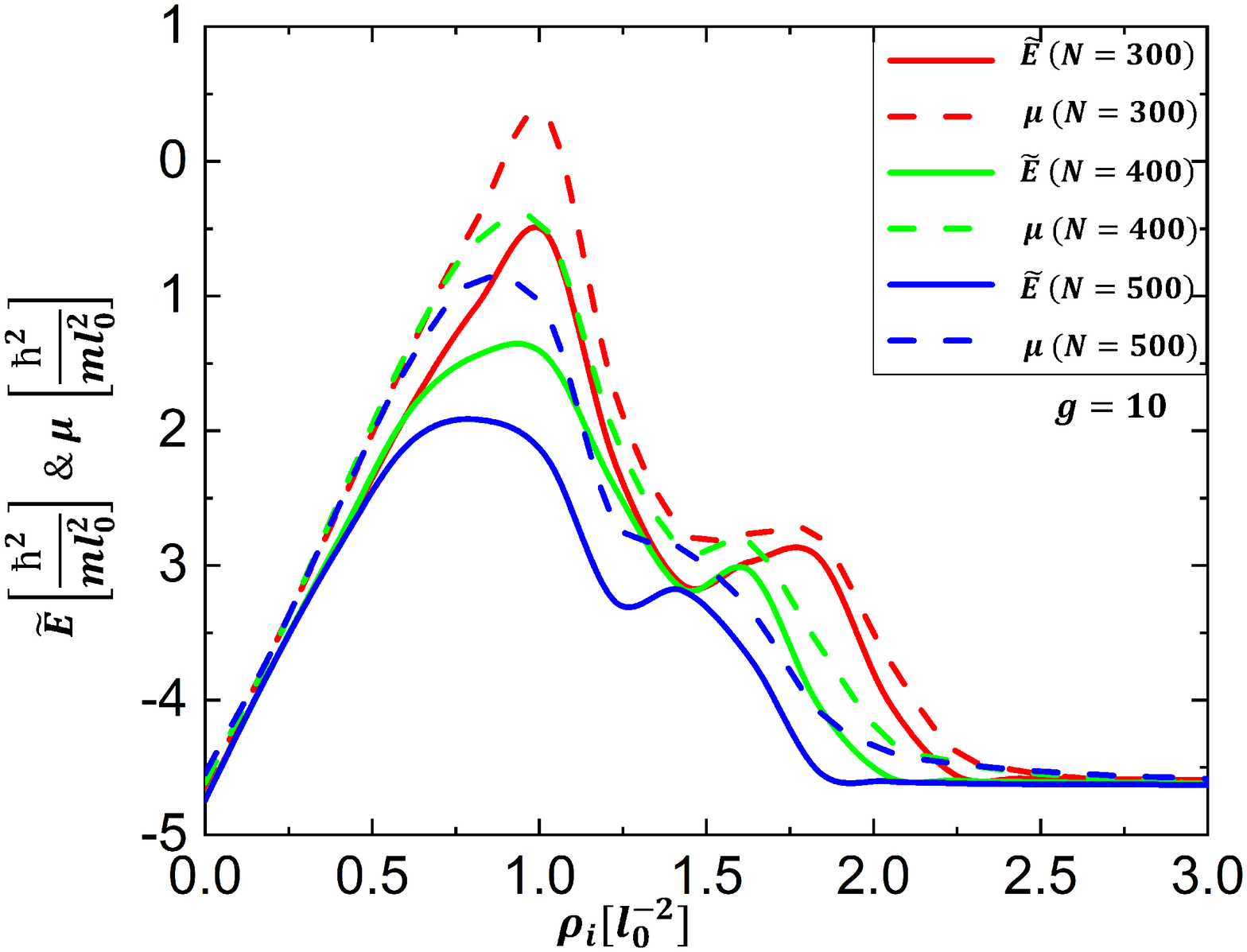}\\
	    \includegraphics[width=9.5cm]{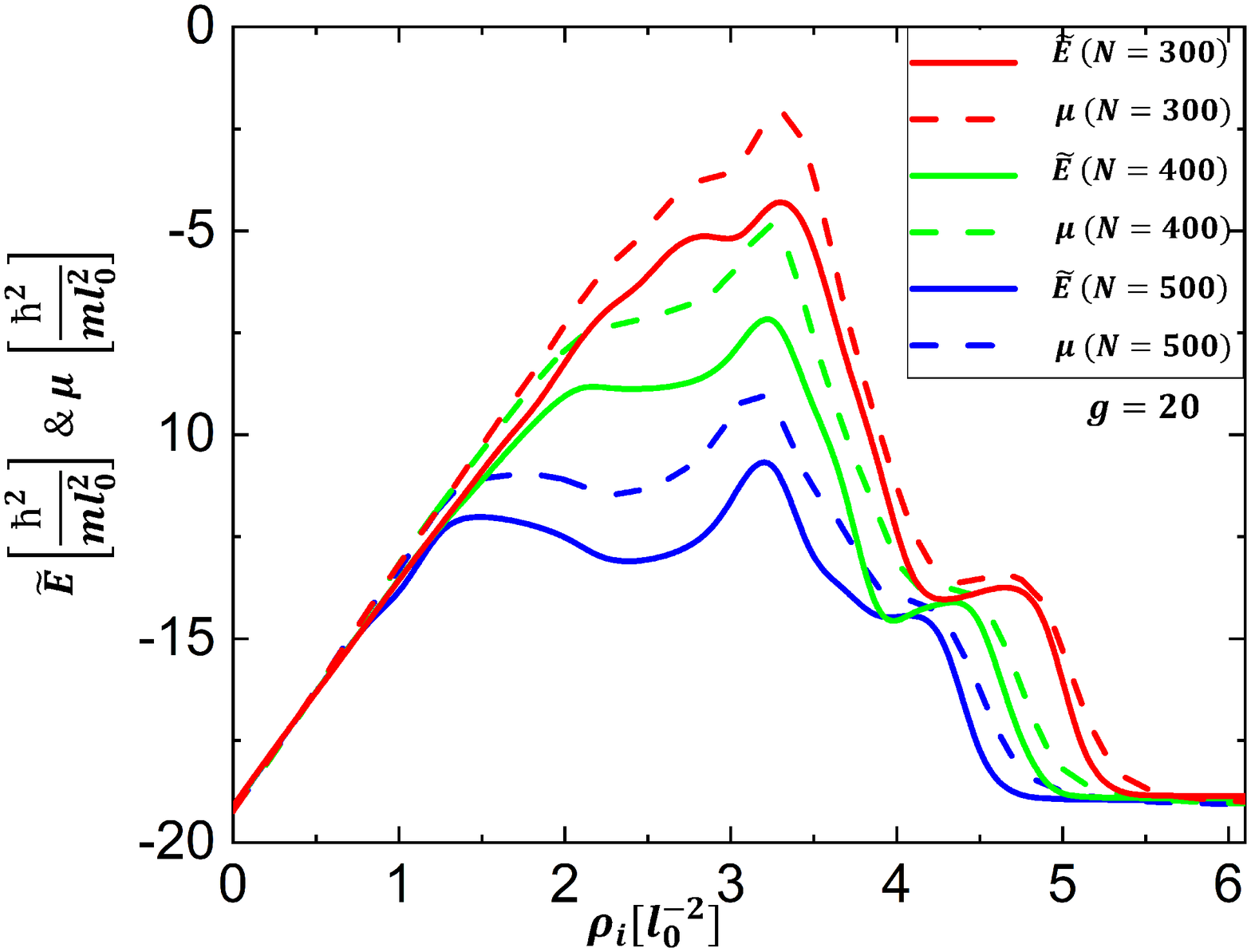}
\caption{ Variation of energy per particle (solid line) and chemical potential (dashed line) with different number of impurity points density for $g = 10$ (upper figure) and $g = 20$ (lower figure). We have included the results for three different particle numbers to see the size dependence nature of the condensate. The error (standard deviation over different realizations of the disorder potential for fixed $\rho_i$) in energy and chemical potential is less than 1\%, which has not been shown here.}
    \label{fig:2Denergy}
  \end{center}
\end{figure}

 After getting the wave function, we have calculated the energy and chemical potential
\begin{eqnarray}
E &=& \int \Big ( \frac{1}{2} \left(|\nabla\psi_1|^2+|\nabla\psi_2|^2 \right) \\
&+& \frac{g}{2} \left( |\psi_1|^2-|\psi_2|^2 \right)^2+ \frac{g^2} {8\pi} \rho^2 \ln (\frac{\rho}{{\sqrt e}}) + V\rho \Big )d^2\vec{r} \nonumber \\
\mu &=& \int \Big ( \frac{1}{2} \left(|\nabla\psi_1|^2+|\nabla\psi_2|^2 \right) \\
&+& g \left( |\psi_1|^2-|\psi_2|^2 \right)^2+ \frac{g^2} {4\pi} \rho^2 \ln \rho + V\rho \Big )d^2\vec{r} \nonumber 
\end{eqnarray}
  
 We have considered twenty different sets of configurations of impurity points to get the disorder average (disorder realization) for a fixed number of impurity point concentration, $\rho_i$.

The disordered average energy of the system is
\begin{eqnarray}
	\tilde{E} = \frac{1}{N_M}\sum^{N_M}_{i=1}   E_i
\end{eqnarray}
where $N_M$ is the total number of impurity configuration (we have considered $N_M=20$) for a fixed number of impurity point density, $E_i$ is the energy for an impurity points distribution.  

% {\color{red} The potential and the density of the condensate has been shown in fig... and fig...}
%Finally we have calculated the Energy and Chemical Potential for different number of impurity points considering different values of $V_0$. \\

\subsection{Results and discussions}

We have calculated the density of particles (solving equation (1)) for interaction strength $g=10$ for different impurities point concentrations, as shown in FIG. \ref{fig:2Dg10den}. From the density profiles of FIG \ref{fig:2Dg10den}  we have noticed as follows:
We have a nice sharp spherical droplet in the absence of the impurity potential (topmost figure of FIG \ref{fig:2Dg10den}). As soon as we introduce the impurity potential, the droplet becomes porous, and if we increase the impurity point density, the droplet starts to segregate. If we further increase the speckle point density, the droplet moves out from the impurity region (last figure of FIG \ref{fig:2Dg10den}) at around 1.9 impurity points per unit area (impurity is there in the region -35 to 35 for x and y-direction). The important point is that the condensate remains in the liquid phase in the presence of large impurity potential.
The behavior of the droplet is similar to the classical droplet of liquid.
%From this study, we can conclude that the liquid phase is more robust than the gas phase of BEC.
The impurity potential and the corresponding density of the condensate for $N=400$ particles have been shown in the FIG \ref{fig:2Dpotl} for visualization.

We have calculated the density of condensate for other systems with different interaction strengths for different impurity concentrations (the result has not been shown here as it is similar to that of the system with $g=10$). 
 The system with $g=20$ tolerates more impurity potential than with $g=10$.

{\bf Variation of Energy with the impurity concentration:} 
We have plotted the energy and chemical potential of the system $g=10$ and $g=20$ for different size of the droplet as a function of impurity point concentration. Energy and chemical potential increases with the increase of the speckle points and attain a maximum, then reduce and become constant as shown in the upper panel of FIG \ref{fig:2Denergy} for the system with $g=10$. The increase of energy and chemical potential is obvious as we have considered repulsive impurity potential, whose contribution to the energy is positive. %, energy is decreasing due to the increase of surface area.
After a particular concentration of impurity points, the energy and the chemical potential started decreasing with the impurity points concentration. 
%This is due to the segregation of the droplet. As we have many droplets, so the total surface area increases due to segregation. In this situation, there will be competition between the surface energy and the energy due to impurity. We have not considered the surface energy\cite{surface_energy} in our calculation, and it is obvious that the total energy will increase as the potential term has a positive contribution.
Due to high impurity potential, the droplet segregates into large number of tiny droplets, and some droplets start to move out of the impurity region; that is why the energy decreases.
 After a certain impurity point concentration, the energy becomes constant as the droplet moves out of the potential region; in this situation, the condensate does not affect by the disorder potential.  

At the large impurity concentration, energy becomes constant, and the value is slightly higher than the energy of the system without impurity. This is because the condensate moves out of the impurity region and is separated into four small droplets, which has a higher total surface area than the original single droplet. If we consider circular impurity region, then we will have ring-shaped droplet just outside of the impurity region (the result has not been shown).

In the absence of impurity, the energy per particle does not depend on the size of the condensate, whereas in the presence, of impurity the energy per particle decreases with the size of the condensate. This suggests that the bigger condensate with higher interaction between atoms tolerates more impurity potential.

%The variation of energy and chemical potential with respect to the impurity point concentration for various impurity strength of the system $g=20$ have been shown in the right panel of FIG  \ref{fig:2Denergy}. We have seen an oscillation, maybe due to the competition between the impurity potential and the surface energy.

%%%%%%%%%%%%%%%%%%%%%%%%%%%%%%%%%%%%%%%%%%%%%
%\section{2D with $\pm$ potential}

\section*{Acknowledgement} We would like to thank Prof. S. Ghosh, IITD for the valuable discussion


\begin{thebibliography}{50}

%%%%%%%%%%%%%%%%%%%%%%%%%%%%%%%%%%%%%%%%%%%%%
\bibitem{Petrov2015} D. S. Petrov, Phys. Rev. Lett. {\bf 115}, 155302 (2015).
\bibitem{Trarruell2018} C. R. Cabrera, L. Tanzi, J. Sanz, B. Naylor, P. Thomas, P. Cheiney, L. Tarruell, Science {\bf 359}, 301 (2018)
\bibitem{drop_exp2} G. Semeghini, G. Ferioli, L. Masi, C. Mazzinghi, L. Wolswijk, F. Minardi, M. Modugno, G. Modugno, M. Inguscio, M. Fattori,  	Phys. Rev. Lett. {\bf 120}, 235301 (2018).
\bibitem{Trarruell2018PRL} P. Cheiney, C. R. Cabrera, J. Sanz, B. Naylor, L. Tanzi, L. Tarruell, Phys. Rev. Lett. {\bf 120}, 135301 (2018). 

\bibitem{dipolar_droplets} Observation of dipolar droplets in 2016: H. Kadau, M. Schmitt, M. Wenzel, C. Wink, T. Maier, I. F. Barbut, and T. Pfau, Nat. Phys. {\bf 530}, 194 (2016); 
I. F. Barbut, H. Kadau, M. Schmitt, M. Wenzel, and T. Pfau, Phys. Rev. Lett. {\bf 116}, 215301 (2016); I. F. Barbut, M. Schmitt, M. Wenzel, H. Kadau, and T. Pfau, J. Phys. B {\bf 49}, 214004 (2016);  M. Schmitt, M. Wenzel, B. Bottcher, I. F. Barbut, and T. Pfau, Nat. Phys. \textbf{539}, 259 (2016);  L. Chomaz, S. Baier, D. Petter, M. J. Mark, F. Wachtler, L.  Santos,  and  F.  Ferlaino,  Phys.  Rev.  X \textbf{6},  041039(2016).
\bibitem{LHY} T. D. Lee and C. N. Yang, Phys. Rev. \textbf{105}, 1119 (1957); T. D. Lee, Kerson Huang, and C. N. Yang, Phys. Rev. \textbf{106}, 1135 (1957).
\bibitem{PRL89}G. Modugno, M. Modugno, F. Riboli, G. Roati, and M. Inguscio, Phys. Rev. Lett. \textbf{89}, 190404 (2002).
\bibitem{PRL100} G. Thalhammer, G. Barontini, L. De Sarlo, J. Catani, F. Minardi and M. Inguscio, Phys. Rev. Lett. \textbf{100}, 210402 (2008).
\bibitem{Itali2020} A. Burchianti, C. D. Errico, M. Prevedelli, F. Ancilotto, M. Modugno, L. Salasnich, F. Minardi and C. Fort, Condens. Matter {\bf 5}, 21 (2020).
\bibitem{spin_BEC} J. Stenger, S. Inouye, D.M. Stamper-Kurn, H.-J. Miesner, A.P. Chikkatur, and W. Ketterle,Nat. Phys. \textbf{396}, 345 (1998); M. S. Chang, C. D. Hamley, M. D. Barrett, J. A. Sauer, K. M. Fortier, W. Zhang, L. You, and M. S. Chapman, Phys. Rev. Lett. \textbf{92}, 140403 (2004).
\bibitem{PRL'101} S. B. Papp, J. M. Pino, and C. E. Wieman, Phys. Rev. Lett. \textbf{101}, 040402 (2008).
\bibitem{2Comp_BEC} Ying-Hai Wu and Jainendra K. Jain, Phys. Rev. B \textbf{87}, 245123 (2013).
\bibitem{2D_liquid} D. S. Petrov and G. E.Astrakharchik, Phys. Rev. Lett. \textbf{117}, 100401 (2016).

\bibitem{2D_liquid1} G. E.Astrakharchik, B. A. Malomed, Phys. Rev. A \textbf{98}, 013631 (2018).
\bibitem{2D_liquid2} P. Zin, M. Pylak, T. Wasak, M. Gajda, Z. Idziaszek, Phys. Rev. A \textbf{98}, 051603(R) (2018).
\bibitem{2D_liquid3} Y. Li, Z. Chen, Z. Luo, C. Huang, H. Tan, W. Pang, B. A. Malomed, Phys. Rev. A {\bf 98}, 063602 (2018).
\bibitem{Gajda19} D. Rakshit, T. Karpiuk, P. Zin, M. Brewczyk, M. Lewenstein, M. Gajda, New J. Phys. \textbf{21}, 073027 (2019).
%%%%%%%%%%%%%%%%%%%%%%%%%%%%%%%%%%%%%%%%%%%%%
\bibitem{SS2020} S. Sahu and D. Majumder, J. Phys. B {\bf 53}, 095301 (2020).
\bibitem{Trivedi1991} W. Krauth, N. Trivedi and D. Ceperley, Phys. Rev. Lett. {\bf 67}, 2307 (1991).
\bibitem{He4porous} J. D. Reeppy,  J. Low Temp. Phys. {\bf 87}, 205(1992); % Superfluid He in porous media
 K. G. Singh and D. S. Rokhsar, Phys. Rev. B {\bf 49}, 9013(1994), % Disordered bosons: condensate and excitations 
G. K. S. Wong, P. A. Crowell, H. A. Cho, and J. D. Reppy, Phys. Rev. B {\bf 48}, 3858(1993); % Superfluid critical behavior in the presence of a dilute correlated impurity
I. F. Herbut, Phys. Rev. B {\bf 61}, 14723(2000). %Critical exponents at the SI transition in dirty-boson system
\bibitem{Huang1992} K. Huang, H. F. Meng, Phys. Rev. Lett. {\bf 69}, 644(1992).
\bibitem{Stringari1994} S. Giorgini, L. Pitaevskii, S. Stringari, Phys. Rev. B {\bf 49}, 12938(1994). 

\bibitem{dynamics_imp1} K. Mukherjee, S. I. Mistakidis, S. Majumder, P. Schmelcher, Phys. Rev. A {\bf 101}, 023615 (2020)
\bibitem{machin_learning} S. Pilati and P. pieri Scientific Reports 9, 5613 (2019).
\bibitem{BBM-2020} A. Boudjemaa and K. Abbas, Phys. Rev. A {\bf 102}, 023325 (2020).

\bibitem{speckle1} J. E. Lye, L. Fallani, M. Modugno, D. S. Wiersma, C. Fort, and M. Inguscio, Phys. Rev. Lett. {\bf 95}, 070401 (2005). % BEC in Random potential, Italy
\bibitem{speckle2} R. C. Kuhn, O. Sigwarth, C. Miniatura, D. Delande and C. A. Muller, New Journal of Phys. {\bf 9}, 161 (2007). 
\bibitem{speckle3} B. Abdullaev and A. Pelster,Eur. Phys. J. D 66, 314 (2012); A. Boudjemaa, Phys. Rev. A 91,
053619 (2015).
\bibitem{speckle4} S. Pilati and P. Pieri , scientific reports, {\bf 9}, 5613 (2019).% machine learling of ultracold atoms with speckle disorder
\bibitem{speckle5} Y. P. Chen, J. Hitchcock, D. Dries, M. Junker, C. Welford, S. E. Pollack, T. A. Corcovilos, R. G. Hulet, Physica D {\bf 238}, 1321(2009).
\bibitem{Gaussian1} M. Kobayashi and M. Tsubota, Phys. Rev. B 66, 174516 (2002); G. M. Falco, A. Pelster and R. Graham,Phys. Rev. A 76, 013624 (2007); C. Krumnow and A. Pelster, Phys.
Rev. A 84, 021608 (2011).
\bibitem{Gaussian_Adhikari} Y. Cheng, S. K. Adhikari, Phys. Rev. A {\bf 82} 013631 (2010).
\bibitem{Gaussian2018} Sh. Mardonov, V. V. Konotop, B. A. Malomed, M. Modugno, and E. Ya. Sherman, Phys. Rev. A {\bf 98}, 023604(2018).% Add SOC: arXiv:1807.06794v2 SOC soliton in random potential, dynamics: 1D pot $V(x) = \sum s_i e^{-(x-x_i)^2/\xi^2}$
\bibitem{GP_random2} E. Akkermans, S. Ghosh and Z. Musslimani, J. Phys. B {\bf 41}, 045302 (2008). % Nume. Study of 1D and interacting BEC in a random potential
\bibitem{Gaussian2014} M. Sajid, I. Ashraf, Laser Phys. {\bf 24}, 115501 (2014). % BEC under Gaussian random potential 
\bibitem{Lorentzian} B. Nikolic, A. Balaz and A. Pelster, Phys. Rev. A 88, 013624 (2013).


\bibitem{Anderson2007} L. Sanchez-Palencia, D. Clement, P. Lugan, P. Bouyer, G. V. Shlyapnikov, and A. Aspect, Phys. Rev. Lett. {\bf 98}, 210401 (2007).
\bibitem{Anderson2010} G. Modugno, Rep. Prog. Phys. {\bf 73}, 102401 (2010). % Anderson localization in BEC
\bibitem{Anderson2008} G. Roati, C. D'Errico, L. Fallani, M. Fattori, C. Fort, M. Zaccanti, G. Modugno, M. Modugno, M. Inguscio, Nature {\bf 453}, 895 (2008) % Anderson localization of a non-interacting BEC
\bibitem{Anderson2021} M. C. P. dos Santos and W. B. Cardoso, Phys. Rev. E {\bf 103}, 052210 (2021)



\bibitem{SIT1_2014} C. D Errico, E. Lucioni, L. Tanzi, L. Gori, G. Roux, I. P. McCulloch, T. Giamarchi, M. Inguscio, and G. Modugno, Phys. Rev. Lett. {\bf 113}, 095301 (2014). % Observation of a disordered bosonic insulator from weak to strong interactions
\bibitem{SIT2_2013} L. Tanzi, E. Lucioni, S. Chaudhuri, L. Gori, A. Kumar, C. D’Errico, M. Inguscio and G. Modugno, Phys. Rev. Lett. {\bf 111}, 115301 (2013). % Transport of Bose gas in 1D disordered lattices at the fluid-insulator transition
\bibitem{SIT3_2004} E. Altman, Y. Kafri, A. Polkovnikov, and G. Refael, Phys. Rev. Lett. {\bf 93}, 150402(2004). % Phase transition in a system of 1D boson with strong disorder
\bibitem{boundImp1} P. Horak, J. Y. Courtois and G. Grynberg,  Phys. Rev. A {\bf 58}, 3953 (1998). %; Atom cooling and trapping by disorder.
\bibitem{PIM_2010_Pilati} S Pilati, S Giorgini, M Modugno and N Prokofev, New J. of Phy. 12, 073003 (2010). % Dilute Bose gas with correlated disorder: Pathintegral MC study.


\bibitem{diff_mc} D. M. Ceperley, Rev. Mod. Phys. {\bf 67}, 279 (1995); G. E. Astrakharchik, J. Boronat,J. Casulleras,and S. Giorgini, Phys. Rev. A. {\bf 66}, 023603(2002)
\bibitem{quantum_mc} Nicolas Laflorencie, Euro.Phys. Lett. {\bf 99}, 66001(2012)
\bibitem{GP_random1} B. Min, T. Li, M. Rosenkranz and W. Bao, Phys. Rev. A {\bf 86}, 053612 (2012); % Subdiffusive spreading of BEC in random potentials
 X. Antoine, W. Bao, C. Besse, Computer Phys. Communication {\bf 184}, 2621 (2013). %  Comp. method for dynamics of nonlinear Schrodinger/GP equation

\bibitem{GP_random3} R. Acosta-Diaz, G. Krein, A. Saldivar, N. F. Svaiter and C. A. D. Zarro, J. Phys. A: Math. Thro. {\bf 52}, 445401 (2019). % Disordered BEC in hard walls trap


\bibitem{analytical_ran1} P. Lugan, D. Clement, P. Bouyer, A. Aspect, and L. Sanchez-Palencia, Phys. Rev. Lett. {\bf 99}, 180402(2007). % Anderson localization of Bogolyubov quasiparticle in interacting BEC




\bibitem{3body2} S. K. Adhikari, Phys. Rev. A {\bf 95}, 023606 (2017).


%\bibitem{} 2019: Disordered BEC in hard wall's trap; calculate quenched free energy using second quantization. (average over the ensemble of all realizations of the quenched disorder)
%\bibitem{} 2011: Bogoliubov excitations of disordered BEC, Gaul Phys. Rev. A 83, 063629 (2011), 2nd quantization.
%\bibitem{} BEC in random potentials, C. R. Physique 5, 129(2004)
\bibitem{MFT1}  V. I. Yukalov and R. Graham, Phys. Rev. A {\bf 75}, 023619 (2007). % BEC systems in random potential: A self-consistance mean field theory.





\bibitem{dimension} S Gautam, A K Adhikari, J. Phys. B {\bf 52}, 055302 (2019); Annals of Phys. {\bf 409}, 167917 (2019).
%\bibitem{speckle3} Supervised machine learning of ultracold atoms with speckle disorder,     S. Pilati and P. Pieri, Scientific Reports volume 9, Article number: 5613 (2019).

%\bibitem{Pfau_disorder1} A fermionic impurity in a dipolar quantum droplet. M. Wenzel, T. Pfau, I F Barbut, Physica Scripta {\bf 93}, 104004(2018).
%\bibitem{droplet_1d} G. E. Astrakharchik and B. A. Malomed, Phys. Rev. A 98, 013631(2018)
\bibitem{1d_soliton} A. Tononi, Y. Wang and L. Salasnich, Phys. Rev. A {\bf 99}, 063618(2019).

\bibitem{ADC_method} K. Kasamatsu, M. Tsubota and M. Ueda, Phys. Rev. A \textbf{67}, 033610 (2003), W. H. Press et al., \textit{Numerical Recipes in C}, Cambridge University Press (1988).

\bibitem{surface_energy} A. Sartori, J. Marino, S. Stringari and A. Recati, New J. Phys. {\bf 17}, 093036(2015).

\end{thebibliography}
\end{document}